\documentclass{article}
\usepackage[utf8]{inputenc} % allow utf-8 input
\usepackage[T1]{fontenc}    % use 8-bit T1 fonts
\usepackage{hyperref}       % hyperlinks
\usepackage{url}            % simple URL typesetting
\usepackage{booktabs}       % professional-quality tables
\usepackage{amsfonts}       % blackboard math symbols
\usepackage{nicefrac}       % compact symbols for 1/2, etc.
\usepackage{microtype}      % microtypography
\usepackage{xcolor}         % colors
\usepackage{comment,amsmath,amssymb,subcaption,float,graphicx,spconf}
\hypersetup{colorlinks=true}
\usepackage[para,online,flushleft]{threeparttable}
\usepackage{multirow}
\usepackage{algpseudocode}
\usepackage{algorithm}
\newcommand\etal{\textit{et al.}}

\newlength\myindent
\title{USM-Lite: Quantization and Sparsity Aware Fine-tuning for Speech Recognition with Universal Speech Models}
\name{\parbox{\linewidth}{\centering Shaojin Ding, David Qiu, David Rim, Yanzhang He, Oleg Rybakov, Bo Li, Rohit Prabhavalkar, \\
\textit{Weiran Wang, Tara N. Sainath, Zhonglin Han, Jian Li$^{1}$, Amir Yazdanbakhsh$^{1*}$, Shivani Agrawal$^{*}$ \thanks{* Equal advising.} }}}
\address{Google LLC, $^{1}$Google DeepMind}
\begin{document}
\ninept
\maketitle
\begin{abstract}
End-to-end automatic speech recognition (ASR) models have seen revolutionary quality gains with the recent development of large-scale universal speech models (USM). However, deploying these massive USMs is extremely expensive due to the enormous memory usage and computational cost. Therefore, model compression is an important research topic to fit USM-based ASR under budget in real-world scenarios. In this study, we propose a USM fine-tuning approach for ASR, with a low-bit quantization and $N$:$M$ structured sparsity aware paradigm on the model weights, reducing the model complexity from parameter precision and matrix topology perspectives. We conducted extensive experiments with a 2-billion parameter USM on a large-scale voice search dataset to evaluate our proposed method. A series of ablation studies validate the effectiveness of up to int4 quantization and 2:4 sparsity. However, a single compression technique fails to recover the performance well under extreme setups including int2 quantization and 1:4 sparsity. By contrast, our proposed method can compress the model to have 9.4\% of the size, at the cost of only 7.3\% relative word error rate (WER) regressions. We also provided in-depth analyses on the results and discussions on the limitations and potential solutions, which would be valuable for future studies.

\end{abstract}
\section{Introduction}
\label{sec:intro}

End-to-end Automatic speech recognition~\cite{wang2019overview, hannun2014deep, graves2012sequence, chorowski2015attention, dong2018speech}, a technique that transcribes audio to text, has been widely integrated into modern user-interactive AI services and devices (e.g., search by voice, voice assistant, etc.). Over the past few years, ASR models have seen quality and latency improvements under diverse test conditions~\cite{li2020comparison,he2019streaming,CC18,KimHoriWatanabe17}. Meanwhile, end-to-end ASR has been shown to dramatically benefit from self-supervised learned (SSL) speech representations~\cite{schneider2019wav2vec, baevski2020wav2vec, hsu2021hubert, chiu2022self} in both quality and production perspectives. Fine-tuning from these self-supervised speech models significantly improves ASR quality. More significantly, they provide a suitable initialization and reduce training cost for all the downstream tasks. 

More recently, with the rapid emergence of high capacity hardware and the availability of large-scale datasets, SSL speech models see a trend of growing larger~\cite{radford2023robust, zhang2023google, pratap2023scaling}. These models scale conventional SSL speech models up, to capture multi-domain and multi-lingual distributions. With such capability, they can serve as a universal foundation model for most of the speech processing tasks. However, the massive size of these models (several billions of parameters) makes them extremely expensive in deployment due to the need of the considerable amount of memory and computational units.
Therefore, efficient fine-tuning and model compression algorithms have become unprecedentedly important research topics.

From prior studies, we have seen success on end-to-end ASR compression through sparse network pruning~\cite{takeda2017node, shangguan2019optimizing, gao2020rethinking, lai2021parp, ding2021audio} and model quantization~\cite{fasoli20214, bie2019simplified, ding20224, rybakov20232}
However, compressing these massive universal speech models can lead to new challenges on the top of regular end-to-end models. For example, USMs have much large model sizes, and therefore higher compression ratios are needed to reach the efficiency requirements for deployments. More importantly, most current compression methods have considerable quality regressions at high compression ratios (e.g., >75\% sparsity, 2-bit quantization), which could lead to inferior user experiences when deployed to productions.

This motives us to investigate the effectiveness of compressing the model from different perspectives at the same time: quantization reduces the model complexity from the parameter precision, while sparsity focusing on the matrix topology. Accordingly, we propose a USM fine-tuning approach for ASR, with a low-bit quantization and $N$:$M$ structured sparsity~\cite{zhou2021learning} aware paradigm on model weights, where both techniques are hardware friendly and are supported by modern GPUs and TPUs. During forward-propagation, we first perform magnitude pruning on a weight matrix with $N$:$M$ sparsity, and then, we perform per-channel quantization on the non-zero weights. At back-propagation time, we adopt straight through estimator (STE)~\cite{bengio2013estimating} to bypass the quantization rounding functions.

We conducted extensive experiments on a large-scale voice search dataset with the 2-billion parameter CTC USM backbone from~\cite{zhang2023google} to evaluate the proposed approach. With our proposed approach of 4-bit quantization and 2:4 sparsity aware training scheme, the model size can be reduced to 9.4\% of the float32 size, at the cost of only 7.3\% of relative WER regressions. Additionally, we systematically benchmark and analyze the effectiveness of quantization and sparsity through an ablation study, and we discuss the results with related studies.
These benchmarks and discussions provide useful information for future research and productions.

\vspace{-3pt}

\section{Related Work}
\label{sec:related_work}

\vspace{-2pt}

\textbf{Self-supervised learning for ASR.} Schneider~\etal~\cite{schneider2019wav2vec} presented the first exemplar study on SSL speech model, which directly learns speech representations from the raw waveform through a contrastive loss. Subsequent studies improves SSL speech models through advanced learning paradigms~\cite{baevski2020wav2vec, hsu2021hubert, chiu2022self} as well as data and model size expansions~\cite{radford2023robust, zhang2023google, pratap2023scaling}. Initializing from SSL pre-trained encoders has significantly improved state-of-the-art ASR performance, especially under a low-supervised-data setup.

\noindent
\textbf{ASR model compression through quantization and sparsity.} A standard design approach to fit the ASR model under budget is to apply model quantization or network pruning to large models. Recent quantization studies~\cite{fasoli20214, bie2019simplified, ding20224, rybakov20232} have shown that it is possible to quantize the ASR models to 4-bit and even 2-bit with only marginal performance loss. Similarly, in terms of network pruning, both unstructured and structured sparsity~\cite{takeda2017node, shangguan2019optimizing, gao2020rethinking, lai2021parp, ding2021audio} patterns have seen reasonable performance at high sparsity level, through various algorithms based on iterative magnitude pruning.

\noindent
\textbf{Relation to prior work.} There are several previous studies that investigate SSL speech model compression~\cite{peng2023dphubert, lai2021parp, lee2022fithubert, jang2023recycle} through sparsity, knowledge distillation, attention re-use, or their combinations. Our proposed study differs from them in several aspects. First, we explore the combination of sparsity and quantization for compression, which has not been investigated in previous studies. The native operations in quantization provides significant speed up during both training and inference time compared to prior studies. More importantly, all of these prior studies show considerable quality drop at high compression ratio. By contrast, our model has only 7.3\% relative WER regression at 9.4\% of the original model size. In addition, we focus on model compression during the ASR fine-tuning stage of SSL models, instead of pre-training. The dense model and compressed model usually have very different distributions at convergence. When initializing from the near-optimal pre-trained weights of the dense model, we need to adapt the distribution to be optimal to the compressed model within limited training steps, which makes it a more challenging task. 
Lastly, the backbone size (2B vs. 10-100M parameters) and data size ($\sim$1M hours vs. $\sim$1,000 hours) are much larger than previous studies.

\section{Method}
\label{sec:method}

\subsection{USM Backbone Architecture}
\label{sec: backbone}

We use the state-of-the-art USM-CTC~\cite{zhang2023google} backbone with 2B parameters in this study. The encoder is comprised of 32 Conformer layers~\cite{gulati2020conformer}, with a dimensionality of 1536. We use relative attention~\cite{dai2019transformer} in the self-attention layer, with 16 attention heads. The kernel size of the depthwise convolution is set to 5. The model is pre-trained with BEST-RQ~\cite{chiu2022self} on over 12 million hours of speech data in over 500 languages collected from YouTube. BEST-RQ runs in the BERT training fashion, which takes the audio as the input and predicts the masked speech representations. In addition, the left and right attention context per layer is set to 128 frames. 

Given a pre-trained USM, we extract the encoder and pair it with a randomly initialized softmax layer corresponding to the word piece model (WPM). Following this, the model is fine-tuned with the vanilla connectionist temporal classification (CTC)~\cite{graves2006connectionist} loss. Although RNN-T~\cite{graves2012sequence} or LAS~\cite{chan2016listen} may lead to improved WER due to the additional language modeling capability, these models runs in an auto-regressive manner during inference, which is hard to parallelize during inference and has a much higher latency with such a massive encoder. By contrast, CTC based model has an encoder only architecture without any auto-regressive dependency that can be easily parallelized, and it is therefore more efficient for large-scale models.

\subsection{Native Quantization Aware Training (QAT)}
\label{sec: qat}

Suppose we have a single linear layer matrix multiplication $\mathbf{Y = X\otimes W}$, where $\mathbf{X}^T\in\mathbb{R}^I$, $\mathbf{Y}^T \in\mathbb{R}^J$
and $\mathbf{W}\in\mathbb{R}^{I\times J}$ are the input, output, and weight, respectively. Running a matrix multiplication with per-channel weight quantization can be represented as:

\begin{equation}
\label{Eq: per channel}
    \mathbf{Y}_j = \mathbf{s}_j  \cdot \left[\mathbf{X} \otimes \text{Quantize}(\mathbf{W}_j)  \right], 1 \leq j \leq J,
\end{equation}

\noindent
where the $\text{Quantize}(\cdot)$ operation does:

\begin{equation}
\label{Eq: quantize}
    \text{Quantize}(\mathbf{W}_j) = \text{round} \left(\frac{\mathbf{W}_{j}}{\mathbf{s}_j} \right),
\end{equation}

\noindent
$\mathbf{s}_j\in\mathbb{R}$ denotes the scale of the $j$-th channel, and $\mathbf{W}_{j}$ is the $j$-th column of $\mathbf{W}$. The scale is computed by dividing the maximum value of $\mathbf{W}_{j}$
with the maximum value of the integer range. For int8 and int4 quantization, we use the simplest symmetric quantization introduced in~\cite{ding20224} (e.g., [-127, 127] for int8 and [-7, 7] for int4).
However, when the precision reduces to 2-bit, symmetric quantization under-utilizes the quantization buckets (i.e., only three values are used). Therefore, following~\cite{rybakov20232}, we adopt asymmetric quantization for int2 models, along with sub-channel quantization, which splits a channel into several groups with dedicated scales for each group.

During the forward propagation of quantization aware training, we apply eq.(\ref{Eq: per channel}) to all the fully-connected layers of the model, and use STE to bypass the rounding function that is not derivative at back propagation (zero derivative almost everywhere).
More importantly, we cast the quantized weight from eq.(\ref{Eq: quantize}) to the \textit{native} integer type. Compared to commonly used ``fake" quantization~\cite{jacob2018quantization} that uses float operations during training and integer operations during inference, this avoids the numerical difference caused by the operation mismatch between training and inference.

\subsection{Magnitude based Pruning with $N$:$M$ Sparsity}
\label{sec: sparsity}

\begin{figure}[t]
    \centering
    \includegraphics[width=\columnwidth]{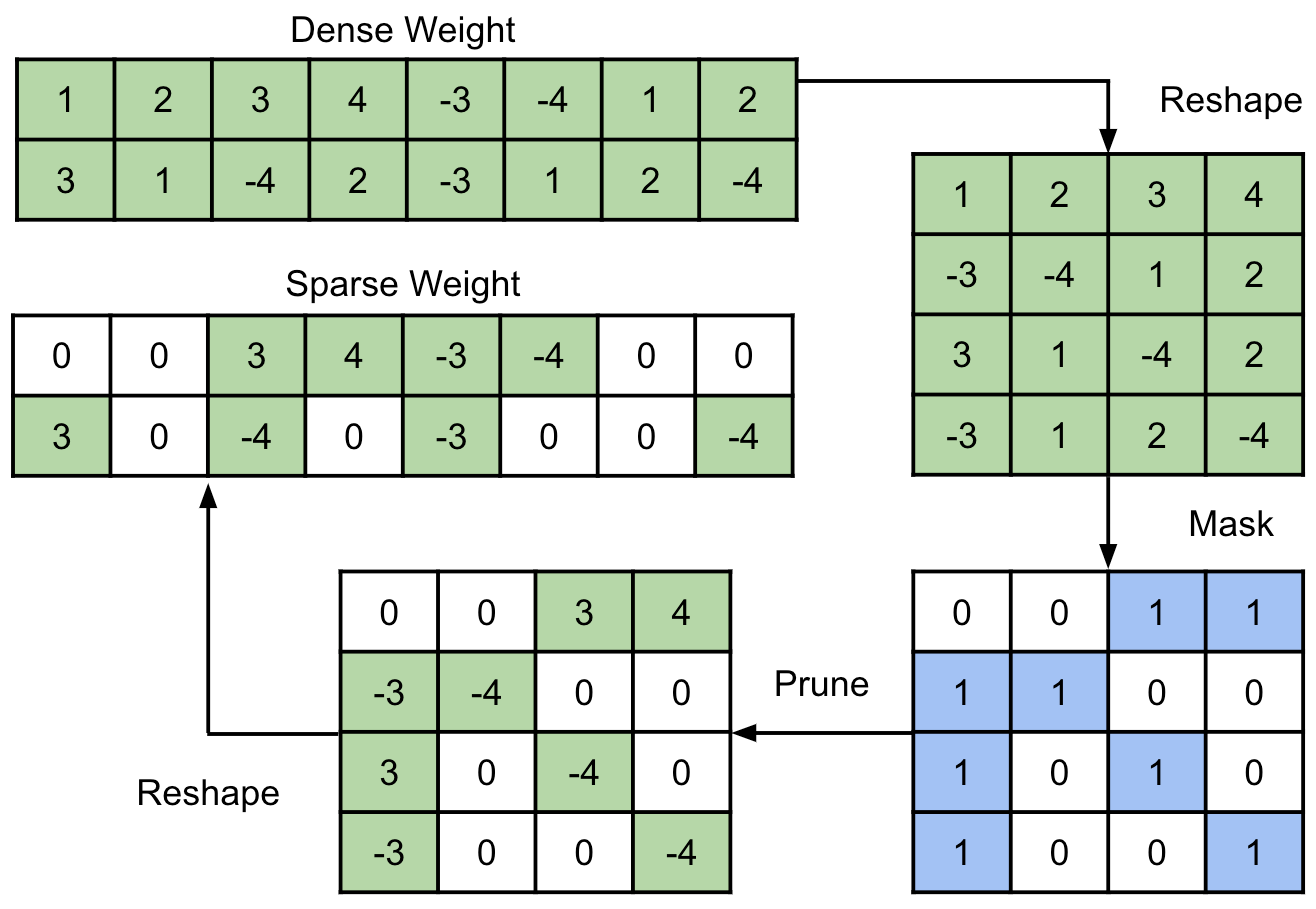}
    \vspace{-20pt}
    \caption{Illustration of magnitude based pruning with $N$:$M$ sparsity on a weight matrix. This example has $N=2$ and $M=4$.}
    \label{fig:n_in_m}
    \vspace{-10pt}
\end{figure}

$N$:$M$ sparsity has a pattern of: for each group of $M$ consecutive weights, there are at most $N$ non-zero values. In this work, we focus on the commonly used case of $M=4$, but it can be easily extended to patterns with arbitrary $M$. As shown in Figure~\ref{fig:n_in_m}, at a pruning step, we first reshape the dense weight matrix to be $\mathbf{V}\in \mathbb{R}^{K\times M} = \text{Reshape}(\mathbf{W})$, where $K$ is the number of groups. Then, we identify the $N$-th largest magnitude weight $\phi_k$ for each group, and generate the binary mask $\mathbf{M}\in \{0, 1\}^{K\times M}$ by:

\begin{equation}
\label{Eq: mask}
    \mathbf{M}_{km} = \begin{cases} 
      1 & |\mathbf{W}_{km}|\geq \phi_k \\
      0 & |\mathbf{W}_{km}| < \phi_k
   \end{cases}, 1 \leq k \leq K, 1 \leq m \leq M,
\end{equation}

\noindent
The reshaped weight $\mathbf{V}$ is pruned by the mask as:

\begin{equation}
\label{Eq: sparsity}
    \text{Prune}(\mathbf{V}) = \mathbf{V}\odot\mathbf{M}
\end{equation}

\noindent
where $\odot$ denotes the element-wise product. Finally we reshape the pruned sparse weight back to the original shape.

We investigate both \textit{one-shot} and \textit{few-shot} pruning in training. One-shot pruning only update the mask once at the beginning of the USM fine-tuning stage, and then we freeze the mask and tune the weights only. Few-shot pruning updates the mask for $T_p$ times at the beginning, and similarly, the mask is frozen afterwards. We do not enable STE proposed in~\cite{zhou2021learning} in few-shot pruning, and only the unpruned weights are updated at each iteration.

\subsection{Joint optimization with Quantization and Sparsity}

Empirically, simply applying \emph{quantization} or \emph{sparsity} alone with high compression ratio introduces inevitable regressions (e.g., see results for 2-bit quantization and 1:4 sparsity in Section~\ref{sec: ablation_quant} and~\ref{sec: ablation_sparse}). To maximally reduce the model size while retaining the WER, we propose to compress the model from the aspects of parameter precision and matrix topology jointly, with a combination of quantization and sparsity. Our proposed approach is in a prune-and-quantize fashion, which makes it more convenient during implementation. As described in Section~\ref{sec: sparsity}, the pruned weights are set to zero, which directly maps to the zero-point of symmetric quantization and has no effect on calculating the scale. The overall training process is shown in Algorithm~\ref{alg:training}.

\vspace{-10pt}
{\begin{algorithm}[H]
    \small
    \caption{Proposed quantization and sparsity aware training process for USM fine-tuning.}
    \begin{algorithmic}[1]
    \State \textbf{Inputs}: Speech-text pairs ($\mathbf{X}$, $\mathbf{Y}$), USM model $f(\mathbf{X}; \mathbf{W})$, CTC loss $L(\cdot)$, total training steps $T$, total pruning steps $T_p$
    \State \textbf{Initialization}: $\mathbf{W}=\mathbf{W}_0$ from BEST-RQ.
    \For{$t$ in $1, 2, \dots, T$}
        \If{$t<T_p$}
            \State Update masks for each weight matrix of $\mathbf{W}$ with eq.(\ref{Eq: mask})
        \EndIf
        \State Prune each weight matrix of $\mathbf{W}$ through the mask with eq.(\ref{Eq: sparsity})
        \State Quantize each weight matrix of $\mathbf{W}$ with eq.(\ref{Eq: quantize}) and run matrix multiplications with eq.(\ref{Eq: per channel})
        \State Compute CTC loss: $L(f(\mathbf{X}; \mathbf{W}), \mathbf{Y})$
        \State Update $\mathbf{W}$ with gradients $\frac{\partial L}{\partial \mathbf{w}}$
    \EndFor
    \State \textbf{Return}: Trained model $f(\mathbf{X}; \mathbf{W})$
    \end{algorithmic}
    \label{alg:training}
\end{algorithm}}

\section{Experimental setup}
\label{sec:exp}

\subsection{Datasets}

We evaluate the proposed techniques on a large-scale voice search task. The pre-training dataset is described in Section~\ref{sec: backbone}. During fine-tuning, we train the model with an in-house dataset of 1.2-million-hour United States English audio-text pairs from voice search. All data are anonymized, and our data handling abides by \textit{Google AI Principles}~\cite{googleaiprinciples}. A small portion of the dataset is  hand-transcribed, and the rest is pseudo-transcribed with a 600-million-parameter teacher model~\cite{Seong22}. In evaluations, we report the WER on 8,884 anonymized and hand-transcribed utterances representing the voice search traffic.

\subsection{Implementation details}

In addition to the USM backbone architecture introduced in Section~\ref{sec: backbone}, the network takes an input sequence of 128-dimensional log Mel-filterbank energies, extracted from a 32ms window and 10ms shifts. The input features are down-sampled by 4 times through two 2-D convolution layers, and projected to 1,536 dimensions through a fully-connected layer, before feeding to the encoder. Each convolution layer has a $3\times 3$ kernels, $2\times 2$ strides, and 128, 32 channels, respectively.

The proposed method is implemented in Pax\footnote{https://github.com/google/paxml} with the layer library Praxis\footnote{https://github.com/google/praxis. We \textbf{open-sourced} our implementations of quantization and sparsity here.}. We do not quantize and sparsify convolution layers and the final softmax layers, as their parameter counts are much lower. The models are trained on Tensor Processing Unit (TPU) v3-128~\cite{TPU} with an Adam optimize of transformer learning rate schedule~\cite{vaswani2017attention}. The input feature and encoder modules have a base learning rate multiplier of 0.5, and the softmax layer has that of 2.0. The warm up steps are set to 5,000 and 1,500 for these modules respectively. All models are trained with 200,000 steps with a batch size of 2,048. During evaluations, we \textit{do not} include any additional language models for rescoring to better benchmark the effectiveness of the compression techniques.

\section{Experiments}

\begin{table}[t]
\begin{center}
\caption{Results of ablation studies on quantization. \textit{Model Size Ratio} is computed as the ratio of the estimated model size relative to \textit{B0}. PTQ refers to post-training quantization. }
\label{table:quant}
\resizebox{\columnwidth}{!}{
\begin{tabular}{|c|l|cc|}
\hline
\multirow{2}{*}{Exp} & \multicolumn{1}{c|}{\multirow{2}{*}{Model}} & Voice Search & Model Size  \\
 & & WER & Ratio \\
\hline
B0 & float32 dense 2B CTC USM & 4.1 & - \\
\hline
E0 & int8 PTQ & 4.2 & 25.0\% \\
E1 & int8 QAT & 4.2 & 25.0\% \\
\hline
E2 & int4 PTQ & 86.7 & 12.5\% \\
E3 & int4 QAT & 4.3 & 12.5\% \\
\hline
E4 & int2 QAT & 99.9 & 6.3\% \\
E5 & int2 QAT + 16 sub-channel & 45.2 & 7.3\% \\
E6 & int2 QAT + 32 sub-channel & 32.0 & 8.3\% \\
E7 & int2 QAT + 64 sub-channel & 12.3 & 10.4\% \\
\hline
\end{tabular}}
\end{center}
\vspace{-20pt}
\end{table}

We conduct three sets of experiments to evaluate our proposed approach: 1) Ablation studies on quantization; 2) Ablation studies on sparsity; 3) Overall performance of the proposed combinations. Through the ablation studies on quantization and sparsity, we aim to examine and characterize the effectiveness of the two techniques on fine-tuning the USM model. In the evaluation of our proposed combinations, we compare to several dense/compression baselines and show the state-of-the-art quality and size reduction obtained from the proposed approach. Note that the sizes of the quantized/sparsified models are \textit{estimated} without considering additional offset required by hardware, so the actual model size will be slightly larger.

\subsection{Ablation Studies on Quantization}
\label{sec: ablation_quant}

Table~\ref{table:quant} shows the results of the ablation studies on quantization with int8, int4, and int2. Experiment \textit{B0} has the performance of the float32 dense 2B-parameter CTC USM, which serves as an upper-bound for both ablation studies. Besides QAT, we also explore the performance of post-training quantization (PTQ) with int8 and int4. As shown from the results, int8 PTQ (\textit{E0}) and QAT (\textit{E1}) can both retain the float32 model's WER. With int4 precision, PTQ (\textit{E2}) leads to much more significantly regressions, while QAT (\textit{E3}) only having marginal WER regressions to the float32 model (4.3 vs. 4.1). When it comes to int2, the extreme quantization setup, vanilla asymmetric quantization does not result in a reasonable WER. With the increasing number of sub-channels, the WER can be gradually improved down to 12.3 at 64 sub-channels (\textit{E7}). However, sub-channel scales introduce extra parameters (i.e., a \textit{float32} parameter per sub-channel), which are equivalent to 1\%, 2\%, and 4.1\% of the original model size on 16, 32, and 64 sub-channel models, respectively. These observations mostly correspond to the conclusions from~\cite{ding20224, rybakov20232}. 

\subsection{Ablation Studies on Sparsity}
\label{sec: ablation_sparse}

Similarly, we show the ablation study results on sparsity in Table~\ref{table:sparsity}. Specifically, we investigate 2:4 and 1:4 structured sparsity patterns along with one-shot and few(1k)-shot pruning schedules. Sparse models need extra 1-bit (i.e., binary) parameters to store masks, which counts to 3.1\% of the original model size.
As shown from the table, one-shot (\textit{E8}) and 1k-shot (\textit{E9}) pruning with 2:4 sparsity have marginal regressions compared to the dense models, with 53.1\% of the original model size. When pruning the model more aggressively with 1:4 sparsity, neither of the two models can retain the dense WER, and at the same time, 1k-shot (\textit{E11}) pruning achieves a superior WER than one-shot (\textit{E10}). A possible explanation could be that 1:4 sparsity is more sensitive to the mask, as it is a much hard task than 2:4 sparsity, and therefore, few-shot update can identify a more effective mask in this case.

\begin{table}[t]
\begin{center}
\caption{Results of ablation studies on $N$:$M$ sparsity. \textit{Model Size Ratio} is computed as the ratio of the estimated model size relative to \textit{B0}. }
\label{table:sparsity}
\resizebox{\columnwidth}{!}{
\begin{tabular}{|c|l|cc|}
\hline
\multirow{2}{*}{Exp} & \multicolumn{1}{c|}{\multirow{2}{*}{Model}} & Voice Search & Model Size  \\
 & & WER & Ratio \\
\hline
B0 & float32 dense 2B CTC USM & 4.1 & - \\
\hline
E8 & 2:4 sparsity one-shot & 4.4 & 53.1\% \\
E9 & 2:4 sparsity 1k-shot & 4.3 & 53.1\% \\
\hline
E10 & 1:4 sparsity one-shot & 11.7 & 28.1\% \\
E11 & 1:4 sparsity 1k-shot & 10.6 & 28.1\% \\
\hline
\end{tabular}}
\end{center}
\vspace{-20pt}
\end{table}

\subsection{Overall Performance of Combining Quantization with Sparsity.}

According to the ablation studies from Section~\ref{sec: ablation_quant} and~\ref{sec: ablation_sparse}, we observe that either quantization or sparsity alone fails to have a reasonable WER at the extreme compression ratio. This motives us to investigate combining the two techniques through our proposed training scheme. Additionally, we train the float32 dense USM baselines with 1B (\textit{B1}), 600M (\textit{B2}), and 300M (\textit{B3}) parameters for reference. Table~\ref{table:proposed} shows the results of this experiment. Our two proposed systems with
int4 quantization and 2:4 sparsity achieves 4.4 (\textit{E12}) and 4.5 (\textit{E13}) WER, respectively.
Compared to baseline \textit{B0}, the model size has been reduced by over 10 times (9.4\% of \textit{B0}), with only 7.3\% relative WER regression (4.4 vs. 4.1). Compared to the float32 dense baselines (\textit{B1} to \textit{B3}), the WER of \textit{E12} is even better than the model with 1B parameters, but it only has about one fifth of the size of the 1B model. Finally, when comparing to the quantization (\textit{E7}) or sparsity (\textit{E11}) models at extreme compression rates, our proposed method significantly reduces the WERs of both models to be much closer to the dense model \textit{B0}, while enjoying even smaller model size (9.4\% vs. 10.4\%). Regarding the comparisons between one-shot and few-shot pruning, we do not find significant WER difference with 2:4 sparsity, which is similar to our observations in~\ref{sec: ablation_sparse}.  In summary, these results corroborate with our claim that compressing the model jointly from the parameter precision and the matrix topology aspects are more effective than an individual technique.

\begin{table}[t]
\begin{center}
\caption{Results of the proposed paradigm of combining quantization and $N$:$M$ sparsity. Results on baseline USM with different model sizes are also presented here for comparisons. \textit{Model Size Ratio} is computed as the ratio of the estimated model size relative to \textit{B0}. }
\label{table:proposed}
\resizebox{\columnwidth}{!}{
\begin{tabular}{|c|l|cc|}
\hline
\multirow{2}{*}{Exp} & \multicolumn{1}{c|}{\multirow{2}{*}{Model}} & Voice Search & Model Size  \\
 & & WER & Ratio \\
\hline
B0 & float32 dense 2B CTC USM & 4.1 & - \\
B1 & float32 dense 1B CTC USM & 4.5 & 50.2\% \\
B2 & float32 dense 600M CTC USM & 4.7 & 33.5\% \\
B3 & float32 dense 300M CTC USM & 5.0 & 18.9\% \\
\hline
E7 & int2 QAT + 64 sub-channel & 12.3 & 10.4\% \\
E11 & 1:4 sparsity 1k-shot & 10.6 & 28.1\% \\
\hline
% E12 & int8 QAT + 2:4 sparsity one-shot & 4.5 (TBC) & 15.6\% \\
E12 & int4 QAT + 2:4 sparsity one-shot & 4.4 & 9.4\% \\
E13 & int4 QAT + 2:4 sparsity 1k-shot & 4.5 & 9.4\% \\
% E12 & int4 QAT + 2:4 sparsity 10k-shot &  & 9.4\% \\
\hline
\end{tabular}}
\end{center}
\vspace{-20pt}
\end{table}

\subsection{Limitations and Discussions}

Although we obtain very encouraging results from the aforementioned experiments, we observe several limitations of this work, which we would like to discuss here and further explore in future work. First, we notice that there are more regressions on int4 PTQ and int2 QAT in this work. One potential reasons could be that we do not enable global variational noise (VN)~\cite{graves2011practical} during fine-tuning in this work, as it is not systematically investigated for USM fine-tuning yet. By contrast, \cite{rybakov20232} has it enabled, which has been shown to improve low-bit quantization performance~\cite{qiu2023rand}.
Second, we have not enable STE during model pruning, which can possibly improve the performance of models with $N:M$ sparsity. In future work, we will examine the two approaches along with other techniques that have been shown to effective regarding quantization (e.g., learnable scale, outlier clipping~\cite{esser2019learned}, etc.) and $N$:$M$ sparsity (e.g., SR-STE~\cite{zhou2021learning}), to seek for better quantized and sparse models. With advanced quantization and sparsity models, we will investigate more aggressive combinations such as int2 plus 2:4 to push the model size to the edge.

\section{Conclusions}
\label{sec:conclusions}

USM has significantly improved the quality and simplified the productionization of ASR along with other downstream tasks. In this paper, we for the first time proposed a joint quantization and sparsity aware paradigm for USM-based ASR
fine-tuning. Through extensive experiments and ablation studies on large-scale datasets, we first benchmarked low-bit quantization and $N$:$M$ structured sparsity on USM fine-tuning, validating the effectiveness of int4 QAT and 2:4 sparsity. More importantly, results suggest that our proposed combination of quantization and sparsity can further reduces the models size to 9.4\% of the original model at the cost of marginal performance regressions. We also provided systematical analyses of our results on each of the compression technique and a discussion on the limitation and potential solutions for future investigations.

\section{Acknowledgements}

We would like to thanks Jeremiah Willcock, Emmanuel Guzman, Xingyu Cai, Wonpyo Park, Peiran Li, and Suvinay Subramanian for the discussions and help in this work.

% \vfill\pagebreak

\bibliographystyle{IEEEbib}
\bibliography{strings,refs}

\end{document}